\documentclass[aps, prl, twocolumn, longbibliography,superscriptaddress]{revtex4-2} 

\usepackage{graphicx,color}  
\usepackage{dcolumn}   
\usepackage{bm}        
\usepackage{amssymb}   
\usepackage{textcomp}
\usepackage[section]{placeins}
\usepackage{float}
\usepackage{epsfig,amssymb,amsmath,amsthm,amsfonts,amsbsy,mathrsfs}
\usepackage{eqnarray}
\usepackage{ifthen}
\usepackage{multirow}
\usepackage{array}
\usepackage[displaymath]{lineno}
\makeatletter

\usepackage{subfigure}

\ifthenelse{1>0}{

}
{

}

\ifthenelse{1>0}{
	\newcommand{\remove}[1]{{\color{red}\sout{#1}}}
}
{
	\newcommand{\remove}[1]{}
}

\usepackage[capitalise]{cleveref}

\NewDocumentCommand{\codeword}{v}{ \texttt{#1} }

\begin{document}

\title{Self-organized criticality driven by droplet influx and random fusion}
\author{Bohan Lyu} 
	\affiliation{Peking-Tsinghua Center for Life Sciences, Peking University, Beijing, China}
\author{Jie Lin}
\affiliation{Peking-Tsinghua Center for Life Sciences, Peking University, Beijing, China}
\affiliation{Center for Quantitative Biology, Peking University, Beijing, China}
\affiliation{School of Physics, Peking University, Beijing, China}
\date{\today}

\begin{abstract}
The droplet size distribution typically decays exponentially in solutions formed by liquid-liquid phase separation. Nevertheless, a power-law distribution of nucleoli volumes has been observed in amphibian oocytes, which appears similar to the cluster size distribution in reaction-limited aggregation. In this work, we study the mechanism of power-law distributed droplet sizes and unveil a self-organized criticality driven by droplet influx and random fusion between droplets. Surprisingly, the droplet size dynamics is governed by a similar Smoluchowski equation as the cluster size in aggregation systems. The system reaches a critical state as the area fraction approaches the critical value at which the droplet size has a power-law distribution with a $1.5$ exponent. Furthermore, the system is also spatially scale-free with a divergent correlation length at the critical state, marked by giant droplet-density fluctuations and power-law decay of the pair correlation function.
\end{abstract}

\maketitle

{\it Introduction.---} Critical phenomena are ubiquitous in various physical systems, marked by a divergent correlation length and scale-free distributions \cite{Stanley1971, Hohenberg1977, Ma2018}. The self-organized criticality (SOC) concept has been successfully applied to various non-equilibrium systems \cite{Bak1987, Bak1988, Turcotte1999, Bak2013, Lin2014, Lin2015}, in which a system self-organizes to a critical state where local perturbations trigger responses across all scales. Similarly, fractal structures of clusters and power-law distributions of cluster sizes have been widely observed in the kinetic aggregation of small particles to form large clusters \cite{Schulthess1980, Weitz1984, Weitz1985, Weitz1986, Ball1987}. Hyperuniform patterns in which density fluctuations are suppressed at large length scales \cite{Torquato2018} are also observed in non-equilibrium systems, such as sheared colloidal suspensions \cite{Corte2008, Wilken2020}, the jamming transition \cite{Wilken2021}, and spinodal decomposition \cite{Wilken2023}.

\begin{figure}[hbt!]
	\centering
	\includegraphics[width=0.4\textwidth]{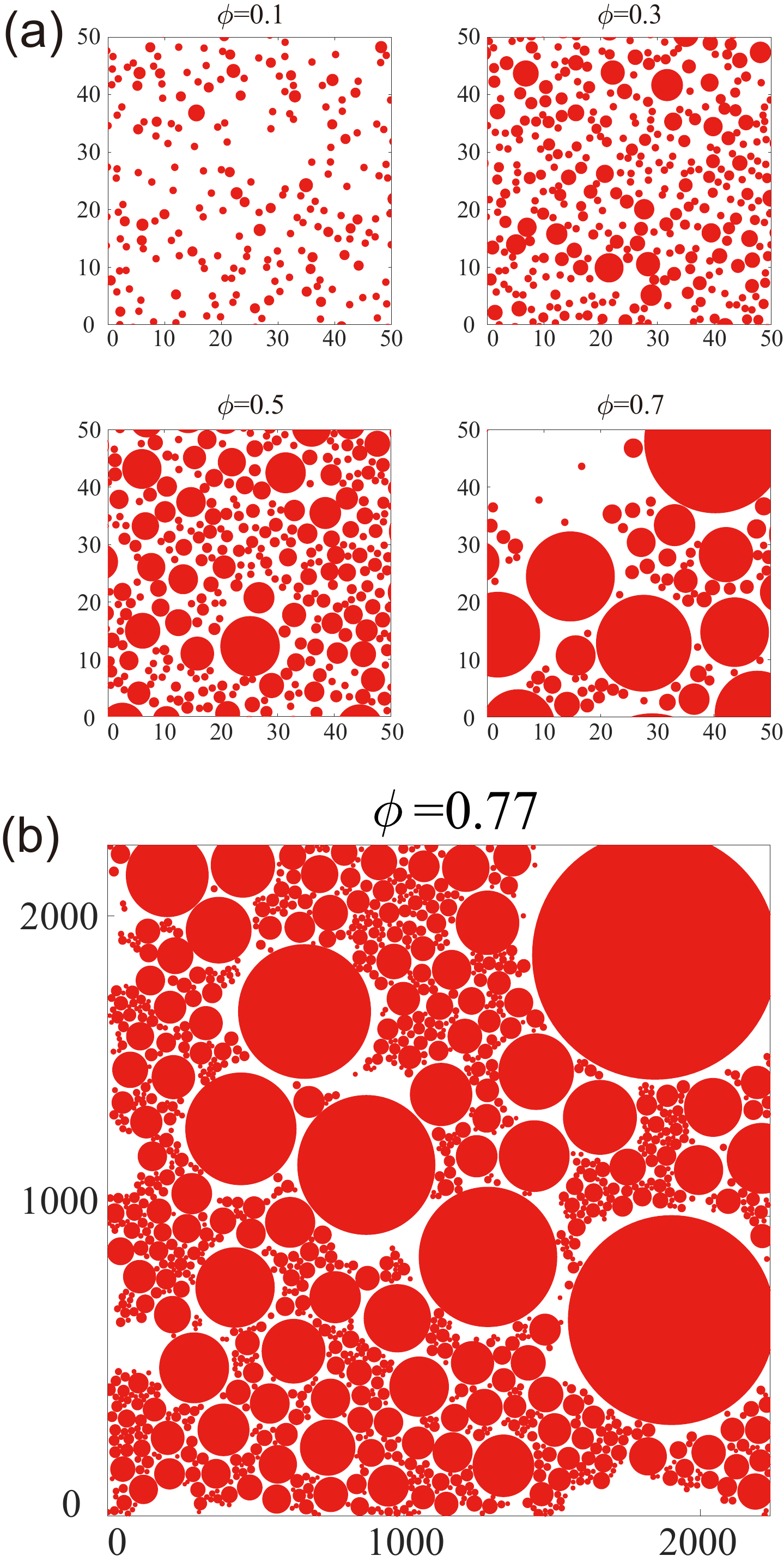}
	\caption{{We randomly add small droplets quasi-statically to the system. As the area fraction of droplets $\phi$ increases, the system approaches a critical state at which scale-free patterns of droplets emerge.}}
	\label{pattern}
\end{figure}

Similar to the power-law distributions of cluster sizes in particle-aggregation systems \cite{Schulthess1980, Weitz1985, Weitz1986, Ball1987}, Brangwynne et al. have observed a power-law distribution of nucleoli volumes in {\it Xenopus laevis} oocytes where the formation of nucleoli is governed by liquid-liquid phase separation \cite{Brangwynne2011}. Power-law distributions of cluster size and nucleoli volume with an exponent $\tau=1.5$ were observed in both cases. However, in the case of aggregation, the power-law distribution happens only in the case of reaction-limited aggregation (RLA) \cite{Weitz1985, Weitz1986}, in which the number of diffusion-induced collisions before aggregation of two clusters succeeds is sufficiently large to allow clusters to sample all possible mutual bonding configuration. This feature of RLA appears inconsistent with nucleoli since these membraneless organelles are liquid-like and fuse virtually immediately once they touch \cite{Brangwynne2011}, a common feature of liquid-liquid phase separation \cite{Brangwynne2009, Hyman2014, Banani2017, Julicher2024}. Therefore, the fusion dynamics of nucleoli seem to be better described by diffusion-limited aggregation (DLA); however, a characteristic size of clusters appears in DLA, and the distribution of cluster sizes typically decays exponentially \cite{Weitz1986, Meakin1990, Berry2018}. Therefore, the mechanism of power-law distributed nucleoli volume observed in {\it X. laevis} oocytes remains elusive. Brangwynne et al. have empirically found that introducing a slow constant influx of small droplets in a Monte Carlo simulation of fusing droplets could numerically reproduce the observed power-law distribution of droplet size, which hints that an external injection may be crucial for generating this power-law distribution of droplet sizes.

In this paper, we introduce a simple model of droplet fusion with a  quasi-static addition of small droplets. We add a spherical droplet of a unit area in a random position to a two-dimensional system with a periodic boundary condition and linear system size $L$. If any droplets overlap, they fuse into a new spherical droplet with a conserved area and a fixed center of mass. We repeat the fusion process until no droplets overlap. Then, we randomly add a new unit-area droplet and repeat the process. We label the system state by the area fraction $\phi$, the total area of all droplets divided by the system area, which gradually increases as more and more droplets are added to the system. Interestingly, the system approaches a critical state as $\phi \rightarrow \phi_c$, where the droplet size distribution becomes scale-free and decays as a power-law (Figure \ref{pattern}), which has not been reported yet as far as we realize. We propose an argument to demonstrate that the fusion kernel, the probability for two droplets to fuse per unit $\phi$, can be approximated as $K(S_1, S_2)\sim S_1+S_2$ and confirm this prediction numerically. Interestingly, this type of kernel leads exactly to an exponent of $1.5$ for the droplet size distribution in the context of kinetic aggregation \cite{Schulthess1980, Dongen1985, Dongen1985a}, in agreement with the distributions of nucleoli volumes \cite{Brangwynne2011}. We also calculate the spatial density variance of droplets inside a finite subsystem and find it decreases slower than the inverse of the subsystem area, which means that the system exhibits a diverging correlation length. Furthermore, the fusion-driven criticality is robust regarding the protocol, and it also emerges in an alternative model in which the area fraction is fixed.

\begin{figure}[htb!]
	\centering
	\includegraphics[width=0.48\textwidth]{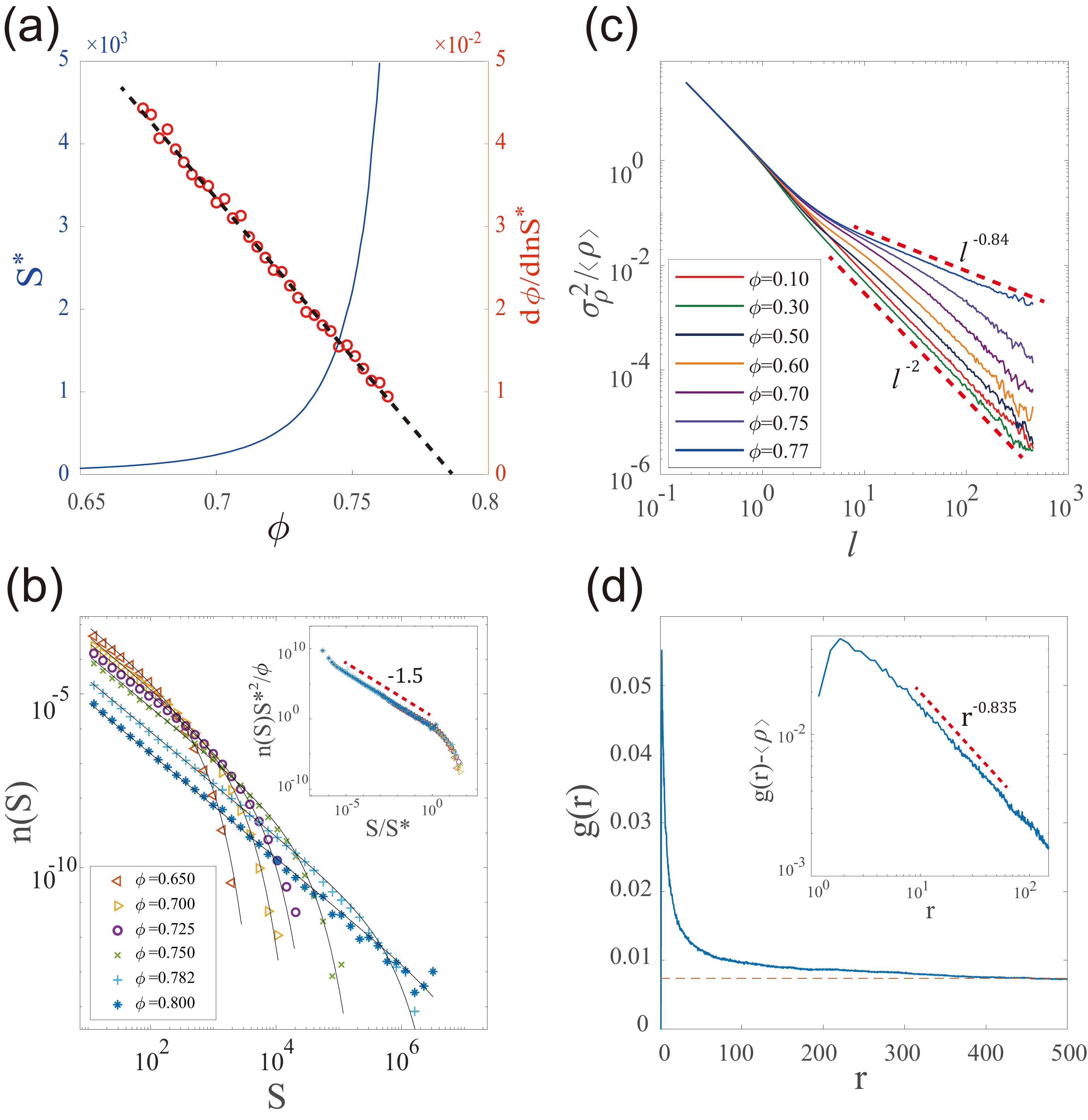}
	\caption{(a) The characteristic droplet size $S^{\ast}$ as a function of area fraction $\phi$. Near the critical point, $S^{\ast}\sim (\phi_c-\phi)^{-\gamma}$, with $\phi_c = 0.782$ and $\gamma = 2.5$.
		(b) Droplet size distributions at various area fractions. The solid lines represent the theoretical prediction Eq. (\ref{size_dirtribution}). The inset shows the normalized distribution, and the dashed line has a slope of $1.5$ in the log-log plot.
		(c) The variance in the spatial density of droplets inside a subsystem as a function of its linear size $l$, normalized by the average spatial density $\langle \rho\rangle$. As $\phi$ approaches the critical value, $\sigma_\rho^2(l) \sim l^{-\eta}$, with $\eta = 0.84$.
		(d) Pair correlation function $g(r)$ at the critical point. The inset shows $g(r)-\langle \rho\rangle$ vs. $r$ in a log-log plot, from which we find that $g(r)-\langle \rho\rangle \sim r^{-\eta}$, with $\eta$ consistent with the value in (c).}
	\label{critical}
\end{figure}

{\it Results.---} To detect the critical state accurately and robustly, we introduce a characteristic droplet size $S^{\ast}$ defined as such: the total area of droplets with $S$ smaller than $S^{\ast}$ is half the total area of all droplets. As we continuously add droplets to the system, $S^{\ast}$ increases as $\phi$ increases and diverges as $\phi\rightarrow \phi_c$ with the power-law scaling
\begin{equation}
	S^{\ast}\sim (\phi_c-\phi)^{-\gamma}.
\end{equation}
The above equation suggests that $d\phi/d\ln S^{\ast}= (\phi_c-\phi)/\gamma$, and we use this linear relationship to extract $\phi_c=0.782$ and $\gamma=2.5$ (Figure \ref{critical}a). 

\begin{figure}[hbt!]
	\centering
	\includegraphics[width=0.47\textwidth]{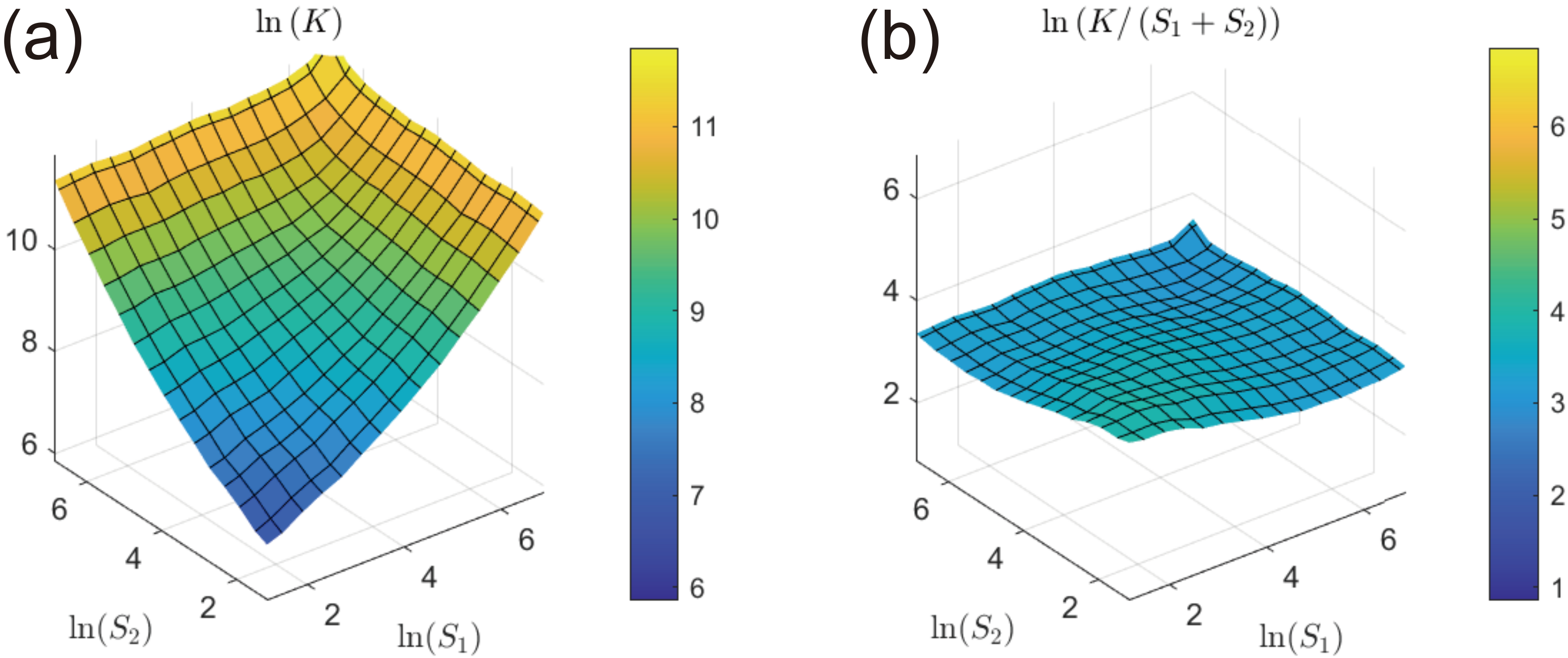}
	\caption{(a) The fusion kernel $K(S_1, S_2)$ in the range $0.7 < \phi < 0.75$. (b) The fusion kernel normalized by $S_1+S_2$ is approximately independent of $S_1$ and $S_2$, in agreement with our prediction that $K(S_1, S_2)\sim S_1+S_2$.}
	\label{kernel}
\end{figure}

We use the Smoluchowski equation to describe the $\phi$-dependence of the droplet size distribution,
\begin{equation}
	\begin{aligned}
		&\frac{dn(S, \phi)}{d\phi} = \frac{1}{2}\sum_{S_1} K(S_1, S-S_1) n(S_1, \phi) n(S-S_1,\phi) \\
		&- n(S,\phi) \sum_{S_1} K(S,S_1) n(S_1, \phi) +\delta_{S, 1},\label{ns}
	\end{aligned}
\end{equation}
where $K(S_1, S_2)$ is the fusion kernel, the probability for two droplets with size $S_1$ and $S_2$ to fuse per unit $\phi$.  Here, $n(S, \phi)$ is the number density of droplet size defined as the number of droplets with size $S$ normalized by the system area $L^2$ at a given $\phi$. In this paper, we interchangeably use the terms number density of droplet size and droplet size distribution. Note that $\phi=\sum_S Sn(S,\phi)$. The last term on the right side of Eq. (\ref{ns}) represents the addition of new droplets with unit area.

In the thermodynamic limit $L\rightarrow \infty$, $n(S, \phi)$ must be system size-independent. Because the dimension of the kernel is length$^{2}$, the only candidate is the droplet size $S$. Another way to think about kernel scaling is to enlarge the system by a factor $\lambda$ to create an imaginary system. For a small interval $\Delta \phi$, the number density of newly created droplets in the original system is given by $\Delta n(S_1 + S_2) = \Delta \phi K(S_1, S_2) n(S_1) n(S_2)$. In the enlarged system, this expression becomes $\Delta n^{\prime}(\lambda^2S_1 + \lambda^2S_2) = \Delta \phi K(\lambda^2 S_1, \lambda^2 S_2) n^{\prime}(\lambda^2 S_1) n^{\prime}(\lambda^2 S_2)$. Since $n^{\prime}(\lambda^2 S) = n(S)/\lambda^2$ by definition, we have $K(\lambda^2 S_1, \lambda^2 S_2) = \lambda^2 K(S_1, S_2)$. Because the two droplets are equivalent, we propose the following form of the fusion kernel, 
\begin{equation}
	K(S_1, S_2) \sim S_1 + S_2,\label{kernel}
\end{equation}
which we verify numerically (Figure \ref{kernel}).


Kernels with such a sum form have an analytical solution in the context of kinetic aggregation of colloidal particles \cite{Schulthess1980}. We propose that the same result should apply to the droplet size distribution, leading to the following asymptotic form of $n(S, \phi)$:
\begin{equation}
	n(S, \phi)= \frac{\phi}{\sqrt{\pi S_c}} S^{-\frac{3}{2}}\exp\Big(-\frac{S}{S_c}\Big),\label{size_dirtribution}
\end{equation}
which satisfies dynamical scaling \cite{Dongen1985, Weitz1986}: 
\begin{equation}
	n(S, \phi) =\phi S_c^{-2} \psi\left(\frac{S}{S_c}\right),\label{collapse_formula}
\end{equation}
where $\psi(x)=x^{-3/2}e^{-x}/\sqrt{\pi}$.

We compute the number densities of droplet size at different \( \phi \), which are well fitted by the theoretical formula Eq. (\ref{size_dirtribution}) with $S_c$ as the fitting parameter (Figure \ref{critical}b). We remark that the characteristic droplet size $S^{\ast}$ is proportional to $S_c$, which can be seen directly from its definition, $\frac{1}{2}=\int_{S^{\ast}}^{\infty} \frac{1}{\sqrt{\pi S_cS} } \exp(-S/S_c)dS = \int_{\frac{S^{\ast}}{S_c}}^{\infty} \frac{1}{\sqrt{\pi}} x^{-1/2}e^{-x}dx$. Therefore, we can replace $S_c$ by $S^{\ast}$  in Eq. (\ref{collapse_formula}), and the dynamical scaling should also work, which we confirm numerically (inset of Figure \ref{critical}b). The power-law exponent below the characteristic cutoff is $1.5$, as predicted. The power-distributed droplet size is also valid for $\phi>\phi_c$. In this case, system-spanning droplets exist, and the droplet size distribution for non-system-spanning droplets follows the power law with the exponent 1.5.

To verify whether there is a diverging correlation length, we randomly draw a square block with a linear size $l$ in the system and compute the spatial density $\rho$ of droplets inside the block (i.e., the number of droplets in the block divided by $l^2$). We repeat this process and calculate the spatial density variance across different blocks. For $\phi$ far below $\phi_c$, the variance $\sigma_{\rho}^2\sim l^{-2}$ according to the central limit theorem \cite{Torquato2018}. Interestingly, near the critical state, the variance decays as a power-law function of $l$ but with a different exponent,
\begin{equation}
	\sigma_{\rho}^2 (l)\sim l^{-\eta}
\end{equation}
where $\eta=0.84$ (Figure \ref{critical}c). We also compute the isotropic pair correlation function $g(r)$, the particle density with a distance $r$ from the particle at the center, using the droplet center at $\phi_c$. One expects that $g(r)$ should decay as a power-law function of the distance with the same exponent $\eta$ 
\begin{equation}
	g(r)-\langle \rho\rangle \sim r^{-\eta}.\label{gr}
\end{equation}
This can be seen from a simple calculation: $\sigma_\rho^2(l)  = \langle \delta N(l)^2\rangle/l^{2d} = \int d\mathbf{r}_1 d\mathbf{r}_2 \langle \delta \rho(\mathbf{r}_1)\delta \rho(\mathbf{r}_2) \rangle /l^{2d} \sim l^{-d} \int (g(r)-\langle\rho\rangle) r^{d-1}dr \sim l^{-\eta}$. Here, $ \delta \rho(\mathbf{r}) = \rho(\mathbf{r})- \langle \rho\rangle$. We confirm Eq. (\ref{gr}) numerically (Figure \ref{critical}d).

\begin{figure}[htb!]
	\centering
	\includegraphics[width=0.45\textwidth]{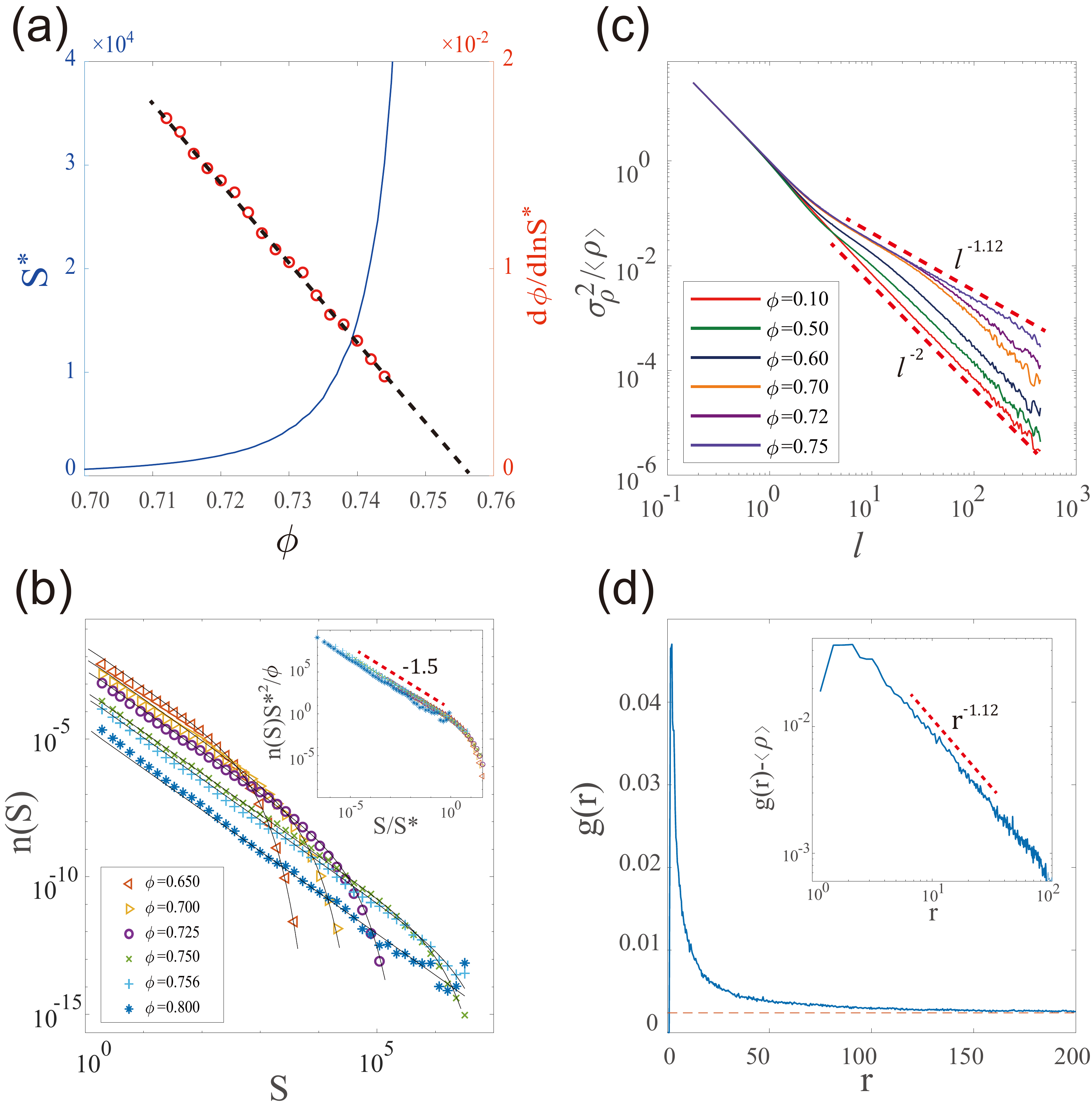}
	\caption{The same analysis as Figure \ref{critical} for the alternative model with a fixed area fraction $\phi$. In this case, $\phi_c = 0.756$, $\gamma = 2.6$, and $\eta = 1.12$, similar to the results of the model in which $\phi$ increases quasi-statically.}
	\label{alternative}
\end{figure}

To test whether the fusion-driven criticality is robust against specific protocols, we also simulate an alternative model in which the area fraction is fixed. Multiple unit-area droplets are simultaneously added to the system with a given area fraction $\phi$. We then let the droplets fuse until there is no overlap between any droplets. In this alternative model, we find essentially the same phenomena. As $\phi$ approaches $\phi_c$, the system reaches a critical state with a diverging correlation length and power-law distributed droplet sizes (Figure \ref{alternative}).

{\it Discussion.---}In this work, we introduce a simple model of droplet fusion where we add small droplets quasi-statically. As the area fraction $\phi$ approaches the critical point, the system becomes scale-free and exhibits power-law distributed droplet sizes. Using an argument based on dimensional analysis, we propose the fusion kernel as a sum form, $K(S_1, S_2)\sim S_1+S_2$. The same argument should hold in three dimensions, where $S$ represents the volume of a droplet. In particle aggregation, the sum-form kernel has been shown to lead to a power-law distributed cluster size with an exponent $1.5$ \cite{Schulthess1980, Weitz1986}. Our findings rationalize the $1.5$ exponent observed in the power-law distribution of nucleoli volume \cite{Brangwynne2011}. Nucleoli function primarily as factories for ribosome subunit biogenesis, and they are RNA/protein bodies within the nucleus \cite{Boisvert2007, Feric2016}. The nucleus of amphibian oocytes is much larger than somatic nuclei; therefore, it is a model system to study the shape and dynamics of nucleoli. We hypothesize that the addition of new nucleoli in amphibian oocytes may be generated by significant rDNA transcription to provide the egg with large numbers of ribosomes.

Interestingly, the number fraction of N-mer in hyperbranched polymers follows a power-law function similar to the number density of droplets in our case \cite{Rubinstein2003}. We argue that in the case of hyperbranched polymers, the probability of two polymers ($N_1$-mer and $N_2$-mer) reacting per unit time is also proportional to the sum of their total number: $N_1 + N_2$. Such a process is described by a sum kernel, leading to the same asymptotic distribution. Power-law distributed bubble sizes have also been observed in motility-induced phase separation \cite{Shi2020}, where new bubbles are nucleated at the expense of the larger ones, in line with the reverse Ostwald scenario \cite{Tjhung2018}. 

The self-organized criticality we uncover for this simple system resembles driven disordered systems such as sheared amorphous solids, in which the systems approach a critical state as the driving force approaches the critical value \cite{Chauve2000, Rosso2003, Rosso2009, Lin2014, Lin2015, Ozawa2018}. In our model, the driving force is the quasi-static addition of new droplets, which drives the system into a novel critical state, which, as far as we know, has yet to be found.

\begin{acknowledgments}
The research was supported by Peking-Tsinghua Center for Life Sciences grants.
\end{acknowledgments}

\bibliography{draft_reference}
\end{document}